\documentclass[twocolumn,showpacs,preprintnumbers,amsmath,amssymb]{revtex4}

\usepackage{graphicx}
\usepackage{dcolumn}
\usepackage{bm}
\usepackage{natbib}
\usepackage{subfigure}  
\newtheorem{Def}{Definition}
\newtheorem{Theorem}{Theorem}
\newtheorem{Cor}{Corollary}
\newcommand{\qed}{\square}

\begin{document}

\title{A measure of centrality based on the spectrum of the Laplacian}

\author{Scott D. Pauls}
\email{scott.d.pauls@dartmouth.edu}
\affiliation{Department of Mathematics, Dartmouth College, Hanover, NH, USA}
\author{Daniel Remondini}
\affiliation{Department of Physics of Bologna University and INFN, Bologna, Italy}

\date{\today}

\begin{abstract}
We introduce a family of new centralities, the $k$-spectral centralities.  $k$-Spectral centrality is a measurement of importance with respect to the deformation of the graph Laplacian associated with the graph.  Due to this connection, $k$-spectral centralities have various interpretations in terms of spectrally determined information.

We explore this centrality in the context of several examples.  While for sparse unweighted networks $1$-spectral centrality behaves similarly to other standard centralities, for dense weighted networks they show different properties. In summary, the $k$-spectral centralities provide a novel and useful measurement of relevance (for single network elements as well as whole subnetworks) distinct from other known measures.
\end{abstract}

\maketitle

\section{Introduction}

Over the last two decades, techniques of network analysis have come to play an important role in the representation and understanding of the structure of complex systems.  In particular,  measures of importance and centrality have allowed to quantify aspects of networks which are fundamental to their structural or dynamical properties.  For example, highly central nodes in social networks can indicate actors with the most influence or, from a different point of view, those actors through which information or disease must pass to reach the full graph.  Different measures are crafted using different mechanisms and assumptions involved in measuring importance.  Betweenness centrality \cite{betw} and related measures \cite{Infc} assign importance to nodes through which many shortest (geodesic) paths pass, while Random Walk Betweenness centrality \cite{newman-rw} considers all possible paths.  Degree centrality simply assigns importance to those nodes with the highest number of edges.  Eigenvector centrality \cite{eig-cen}, on the other hand, measures a type of relative importance - a node is important if it is neighbors with other important nodes (see \cite{newman} for a review and discussion of many centrality measures).  While these measures are often (roughly) commensurate when analyzing sparse symmetric graphs, they can diverge dramatically when considering dense symmetric graphs. The basic reason is straightforward - dense graphs have many short paths as most nodes are connected to one another.  Thus, measures such as betweenness and degree centrality can become less meaningful - if all nodes have roughly the same degree and almost every node is connected to every other node, these measures carry very little information. In this paper, we introduce a new family of measures of centrality, $k$-spectral centralities, which can be more effective for dense graphs.

$k$-Spectral centralities are measures of importance relative to properties of the graph Laplacian, defined as $L = D-A$, in which $A$ is the (symmetric) adjacency matrix and $D$ a diagonal matrix with node connectivities as its terms: $D_{ii}=d_i = \sum_j a_{ij}$.
There are many interpretations of the information encoded in the graph Laplacian (and hence measured by the spectral centralities).  First, the graph Laplacian is known to encode various aspects of the geometry and dynamics of the graph.  For example, the number of zero eigenvalues gives the number of connected components, the size of the first non-trivial eigenvalue -  the Fiedler value - encodes aspects of the graph's connectivity \cite{Chung,F1,F2}.  Alternatively, the Fiedler value encodes the rate at which a network can synchronize \cite{lambda2}.  Second, the Laplacian operator's action on function supported on the graph determines the graph theoretic analogue of heat flow (see \cite{LaplEig}).

The spectral decomposition of the Laplacian therefore encodes the various modes of the heat flow.  So, for networks where such mechanism are appropriate (e.g. in models of the evolution of opinion in social networks) these modes identify relevant substructures of the graph.  Third, the graph Laplacian plays a key role in the graph segmentation - into two pieces, via the relaxed Ratio-Cut problem \cite{rcut}, or into multiple clusters using spectral clustering \cite{specclus}.

With these points of view in mind, we define $k$-spectral centrality of a subset, $\mathcal{B}$, of a connected graph (e.g. a single edge or node, or a whole subnetwork) by measuring the subset's relevance in terms of eigenvalues of the graph Laplacian.  Let $A$ be the adjacency matrix of the graph and $L$ its associated graph Laplacian.  We define the $k$-spectral centrality of $\mathcal{B}$ to be $|\lambda_k'(0)|$ where $\lambda_k(t)$ is the $k^\text{th}$ smallest nontrivial eigenvalue of the graph Laplacian of $A_\mathcal{B}(t)$, the adjacency matrix for the network under a perturbation of $\mathcal{B}$.  While the definition applies to generic subsets, we focus primarily on two special cases - single edges and single nodes (i.e. a node with all of its edges).

$k$-spectral centrality measures the extent to which a subset - e.g. a node or edge - contributes to the information encoded in the $k^\text{th}$ eigenvalue/eigenvector pair.  For example, $1$-spectral centrality encodes information on network connectivity.  So, perturbations of subsets with high $1$-spectral centrality have the most impact on connectivity.  While this is similar in some respects to other centralities, for example betweenness centrality, in many cases $1$-spectral centrality is a distinct measure.  Similarly, higher spectral centralities provide measures of higher order connectivity as demonstrated by their role in efficient decomposition of graphs via spectral clustering.

The computation of $1$-spectral centrality for a network of $N$ of nodes is on the order of $O(N^3)$. This is comparable to computational cost for other centralities such as random walk or eigenvector centrality, but we remark that many optimized algorithms (e.g. for betweenness centrality) consider usually a sparse graph (with number of links of the same order of magnitude of the number of ondes) thus they could be considerably slower on dense graphs.
We examine $1$-spectral centrality and other standard centralities for several sparse networks: a toy model, the Zachary karate club network \cite{karate} and a social network of dolphins \cite{dolphin}.  We then move to larger, dense networks:  a network formed from roll call vote data for the $111^{\text{th}}$ U.S. House of Representatives, a correlation network of equities prices from the S\&P 500 and a biological weighted network obtained from the Immune System \cite{Tieri05}.  Generally, we see that for the small, sparse networks, $1$-spectral centrality behaves similarly to existing centralities while for the denser examples, there are significant differences.  Higher spectral centralities ($k>1$) behave differently than existing centralities in all examples.

We also discuss the spectral centralities in the context of applications to opinion propagation and consensus formation \cite{degroot, Lehrer, OSM, OSFM} where they have direct interpretation.

\section{Spectral Graph Theory}

For this discussion, we assume that $A$ is symmetric, connected, and without self loops, $a_{ii}=0$ (weighted undirected graphs).  As $A$ is symmetric, $L$ is symmetric and diagonalizable with eigendata given by $(\lambda_i,v_i)_{i=0}^n$ with $\lambda_0 \le \lambda_1 \le \lambda_2 \le \dots \le \lambda_n$.  We observe that $\lambda_0=0$ as $L\bf{1} = 0$.  As we assume the network is connected, we know that $\lambda_1 >0$ \cite{Chung,F1,F2}.  The first nontrivial eigenvalue/eigenvector pair, known as the Fiedler value and vector, have been extensively studied (see, for example, \cite{F1,F2} for the initial work).

The graph Laplacian is directly related to the so-called Ratio-Cut problem.  For a graph with vertex set $V$ and edge set $E$, a solution to the ratio-cut problem is a collection of two subsets of nodes $U$ and $W$ so that $V = U \cup W$ and
\[RCut(U,W)= \frac{\#\{e \in E| \exists u \in U, w \in W \text{ with } e=(u,w)\}}{\# U \cdot \# W\}}\] is minimized.  This variant of the standard min-cut problem favors solutions in which the sizes of $U$ and $W$ are comparable.  Solving the ratio cut problem is NP-hard, but one can solve a relaxed version of the problem via an analysis of the graph Laplacian \cite{rcut}.  More precisely, if $v_1$ is the Fiedler vector associated with the graph Laplacian, then $sign(v_1)$ gives an indicator function which provides an approximation of the solution - i.e. $U = \{i|v_1(i)<0\}, V=\{i|v_1(i)>0\}$.
The first $l$ eigenvectors associated with nonzero eigenvalues provide a dimension reduction of the data.  This dimension reduction, the $l$-dimensional {\em spectral embedding}, simply uses the $l$ vectors to give coordinates of the nodes in an $l$ dimensional space.  In our examples, many of the visualizations are created using the 2-dimensional spectral embedding.  The spectral embedding is the basis for the technique of spectral clustering \cite{specclus}, where a Euclidean clustering method such as k-means is used to cluster the nodes given these coordinates.

\subsection*{Spectral Centrality}\label{spec-cen}
With the setup of the previous section, we define spectral centralities in terms of the effect of a deformation on the size of the eigenvalues.
   \begin{Def} For an adjacency matrix $A$ of a graph $\mathcal{G}$ and a deformation of a subset $\mathcal{B}$ given by $A_\mathcal{B}:(-\delta,\delta) \rightarrow \mathcal{M}$, a differentiable map of $A$ in the space of symmetric matrices, the $k$-spectral centrality associated to the deformation is given by
   \[s^k_\mathcal{B}=|\lambda_k'(0)|\]
   where $\lambda_k(\epsilon)$ is a $k^{\text{th}}$ nontrivial eigenvalue of the graph Laplacian of $A_\mathcal{B}(\epsilon)$.
   \end{Def}

Our primary example is $1$-spectral centrality.  As the Fiedler value, $\lambda_1$ defines the so-called {\em algebraic connectivity}, we provide bounds for the vertex and edge connectivities of the graph (\cite{F1,F2}).  Thus, we see from the definition that $1$-spectral centrality is a measure of how much the deformation changes the algebraic connectivity (and hence the standard connectivity) of the network.  As $\lambda_1$ also measures the speed of convergence of the synchronization dynamics on a network \cite{lambda2}, $1$-spectral centrality can also be interpreted in terms of this.  Other $k$-centralities have different interpretations as well. For our purposes, we will mainly focus on $1$-spectral centrality and, in discussing higher spectral centralities, we will do so in the context of the role of the $(\lambda_k,v_k)$ in algorithms such as spectral clustering.  We note that spectral perturbations arise in many applications and that the general idea of measuring importance via perturbation is similar to the to $\delta${\em -centrality} framework of \cite{Infc}.

\begin{Theorem}\label{main}  Let $A_0$ be the adjacency matrix of a connected undirected network with $k$ nodes.  Suppose $A: (-\delta,\delta) \rightarrow \mathcal{M}$ be a differentiable path in the space of symmetric matrices with $A(0)=A_0$ giving a deformation of the matrix $A_0$.  Let $L(\epsilon)$ be the graph Laplacian associated with $A(\epsilon)$ for each $\epsilon \in (-\delta,\delta)$.  Let $(\lambda_i(\epsilon),v_i(\epsilon))$ be the eigenvalue/unit eigenvector pairs for $L(\epsilon)$ and assume that $\lambda_k(0)$ has multiplicity one.   For $\delta$ sufficiently small, $\lambda_k: (-\delta,\delta) \rightarrow \mathbb{R}$ and $v_k: (-\delta,\delta) \rightarrow S^{k-1}$, the eigenvalues and vectors associated with $L(\epsilon)$, are differentiable functions.  Then,
\[\lambda_k'(0)=\sum_{ij} C_{ij}\]
where, $C=FL'(0)F$, and $F$ is the diagonal matrix with entries given by the entries of the eigenvector, $v_k(0)$.
\end{Theorem}
\noindent
{\bf Proof: }
The existence of $\delta>0$ so that $\lambda_k(\epsilon)$ and $v_k(\epsilon)$ are smooth follows from an application of the Implicit Function Theorem to the defining equation of eigenvalues and the defining equation of eigenvectors respectively. Since we assume that network is connected, we know that $\lambda_k(0)>0$ for $k>0$.  This, together with the multiplicity assumption implies that, shrinking $\delta$ if necessary, we may conclude that $(\lambda_k(\epsilon),v_k(\epsilon))$ remain the $k^\text{th}$ nontrivial eigenvalue/eigenvector pair for $\epsilon \in (-\delta,\delta)$.

As $\lambda_k(\epsilon)$ is an eigenvalue of $L(\epsilon)$, we have
\[\det(L(\epsilon)-\lambda_k(\epsilon)I)=0.\]
Differentiating with respect to $\epsilon$ using Jacobi's formula \cite{Jacobi} yields:
\begin{equation*}
\begin{split}
0&=\frac{d}{d\epsilon} \det(L(\epsilon)-\lambda_k(\epsilon)I)\\
&= tr(adj(L(\epsilon)-\lambda_k(\epsilon)I)\cdot (L'(\epsilon)-\lambda_k'(\epsilon)I))
\end{split}
\end{equation*}
where $adj(M)$ is the adjugate of the matrix $M$.  $L(\epsilon)-\lambda_k(\epsilon)I$ is diagonalizable and, as we assume the multiplicity of the $k^\text{th}$ eigenvector of $L(0)$ is one, has a single zero eigenvalue with associated eigenvector $v_k(\epsilon)$.  Letting $V$ be the matrix of orthonormal eigenvectors and $\{0=\mu_1,\mu_2,\dots,\mu_n\}$ be the eigenvalues, we have
\[adj(L(\epsilon)-\lambda_k(\epsilon)I) = det(VV')Vadj(M)V'\]
where $M$ is the diagonal matrix of $\mu_i$. As $V'V=I$, we have $det(VV')=det(V'V)=1$.  The adjugate of the diagonal matrix $M$ is a diagonal matrix $\mathcal{A}$ where $\mathcal{A}_{ii} = \prod_{j\neq i} \mu_j$.  In particular, as $\mu_1=0$, only $\mathcal{A}_{11}$ is nonzero. We denote $\mathcal{A}_{11}$ by $\Lambda(\epsilon)$.  So, in summary we have,
\[adj(L(\epsilon)-\lambda_k(\epsilon)I)= \Lambda(\epsilon) v_k(\epsilon)v_k(\epsilon)^t.\]
Continuing our computation,
\begin{equation*}
\begin{split}
0&=tr(adj(L(\epsilon)-\lambda_k(\epsilon)I)\cdot (L'(\epsilon)-\lambda_k'(\epsilon)I))\\
&= \Lambda(\epsilon)tr(v_k(\epsilon)v_k(\epsilon)^t (L'(\epsilon)-\lambda_k'(\epsilon)I))\\
&= \Lambda(\epsilon)(tr(v_k(\epsilon)v_k(\epsilon)^t L'(\epsilon))-\lambda_k'(\epsilon)tr(v_k(\epsilon)v_k(\epsilon)^t))\\
&= \Lambda(\epsilon)(tr(v_k(\epsilon)v_k(\epsilon)^t L'(\epsilon)))-\Lambda(\epsilon)\lambda_k'(\epsilon)
\end{split}
\end{equation*}
The $tr(v_k(\epsilon)v_k(\epsilon)^t))=1$ since $v_k(\epsilon)$ is taken to have unit length for all $\epsilon$.  Solving for $\lambda_k'(\epsilon)$ yields
\begin{equation*}
\begin{split}
\lambda_k'(\epsilon) &= tr(v_k(\epsilon)v_k(\epsilon)^t L'(\epsilon))\\
&= \sum_{ij} \nu_i(\epsilon) L'(\epsilon) \nu_j(\epsilon)
\end{split}
\end{equation*}
The result follows by letting $C=F(\epsilon)L'(\epsilon)F(\epsilon)$ and evaluating at $\epsilon=0$.  $\qed$

We make two comments.  First, if an eigenvector has multiplicity greater than one, the deformation may still be constructed and analyzed.  However, since there are potentially multiple independent vectors in the kernel of $L(\epsilon)-\lambda_k(\epsilon)I$, the calculation breaks down.  Second, as eigenvectors are not unique, the computation is not {\em a priori} well posed.  However, as we see from the Theorem, $|\lambda_k'(0)|$ is independent of the choice of unit eigenvector.

Next, we apply this to specific deformations using the graph Laplacian $L=D-A$.  Computations using other definitions of the Laplacian and other deformations are easily computed from the previous theorem.

\subsubsection*{Edge centrality}
For an edge between node $i$ and node $j$, we consider the matrix $B^{(i,j)}$ where
\begin{equation*}
B^{(i,j)}_{kl} =
\begin{cases}
1 \;\; \text{if $(k,l)\in \{(i,j),(j,i)\}$}\\
0 \;\; \text{otherwise}
\end{cases}
\end{equation*}
Let $\tilde{L}$ be the graph Laplacian of $B^{(i,j)}$.  Then, observe that $L+\epsilon\tilde{L}$ is the graph Laplacian of the original network under a deformation of the edge between node $i$ and node $j$.  Let $(\lambda_i(\epsilon),v_i(\epsilon))_{i=0}^n$ be the eigendata associated with this deformed Laplacian.
\begin{Cor}\label{edgethm}
 Let $A$ be an adjacency matrix which represents a connected, undirected, weighted network.  For the edge deformation $B^{(i,j)}$,
\[s^k_{(i,j)} = (\nu_i(0)-\nu_j(0))^2\]
where $v_k(0)=(\nu_1(0),\dots,\nu_n(0))$ is the $k^\text{th}$ eigenvector of the graph Laplacian associated with $A$.
\end{Cor}

 Considering the $1$-spectral centrality, this result reflects the intuition given by the link between this graph Laplacian and the ratio-cut problem.  Edges have high $1$-spectral centrality if they connect two nodes which are very far apart when measured by the Fielder vector values.  These are precisely the edges of highest importance to the relaxed ratio-cut problem.
\subsubsection*{Node centrality}

In a similar manner, we define the node spectral centrality by first defining a deformation.  For node $i$, we consider the matrix $B^{(i)}$ where
\begin{equation*}
B^{(i)}_{kl} =
\begin{cases}
A_{kl} \;\; \text{if $(k,l)\in \{(i,\star),(\star,i)\}$}\\
0 \;\; \text{otherwise}
\end{cases}
\end{equation*}
Notice that we have picked the deformation slightly differently, adding the weights of edges.  Letting $\tilde{L}$ be the graph Laplacian of $B^{(i)}$, $L+\epsilon\tilde{L}$ is again the graph Laplacian of the original network under a deformation of all of the edges associated with node $i$ proportionally to their weight.  Note that $\tilde{L}$ has a more complicated form that in the case of a single edge:
\begin{equation}\label{eq1}
\tilde{L}= \begin{pmatrix}
a_{1i} & 0 & \dots & 0 & -a_{1i}& 0 & \dots & 0\\
\vdots &  &  &   &  \vdots&   &       & \vdots\\
-a_{i1} &-a_{i2}& \dots &-a_{i,i-1} & \sum_j a_{ij} & -a_{i,i+1}& \dots &-a_{in}\\
\vdots &  &  &   &  \vdots &    &       & \vdots\\
0 & & \dots & 0 & -a_{ni} & 0 & \dots & a_{in}
\end{pmatrix}
\end{equation}
With this notation in place we have the following result.
\begin{Cor}\label{nodethm}
 Let $A$ be an adjacency matrix which represents a connected, undirected, weighted network.  For the node deformation $B^{(i)}$,

\begin{equation}
s^k_i=\sum_{j=1}^n A_{ij}(\nu_i(0)-\nu_j(0))^2
\end{equation}\label{speccentgen}
where $v_k(0)=(\nu_1(0),\dots,\nu_n(0))$ is the $k^\text{th}$ eigenvector of the graph Laplacian associated with $A$.
\end{Cor}

Similarly to edge $1$-spectral centrality, this makes sense in the context of the relaxed ratio cut problem - a node is as important as the weighted sum of its edges.
\subsubsection*{Spectral centrality for general subsets}
We can also do this with any subgraph - simply pick the deformation matrix, $B$, to be the adjacency matrix of the subgraph in question.  As before, we let $\tilde{L}$ be the graph Laplacian associated to $B$ and we find a similar result.  Specifically, if $A$ be an adjacency matrix which represents a connected, undirected, weighted network.  For the general deformation $B$,
\[s^k_B= \sum_{i,j} B_{ij}(\nu_i(0)-\nu_j(0))^2\]
where $v_k(0)=(\nu_1(0),\dots,\nu_n(0))$ is the $k^\text{th}$ eigenvector of the graph Laplacian associated with $A$.

\subsubsection*{Relation to other measures of centrality}

In \cite{Borgatti}, Borgatti introduced a typology of centrality measures based on characterizations of their flow processes.  The graph Laplacian induces a flow process which, in that terminology, follows {\em walk trajectories} - paths that are traced on the network can loop - and spreads via a {\em parallel process} - nodes spread simultaneously to all neighbors rather than just one.  To make this precise, we describe the flow process induced by the graph Laplacian in terms of information propagation, where we view the information on the network as a function supported on the nodes.  The values of this function then change over time according to the flow dynamics.  In the case of dynamics governed by the graph Laplacian, the function value at the node is replaced by a value proportional to the weighted average of the function values over all the node's neighbors:
\[ x_{t+1}= x_t-Lx_t = x_t-Dx_t+Ax_t\]
so $x_{t+1}(i)=x_t(i)-\sum_j A_{ij} (x_t(i)- x_t(j))$. This is the discrete analogue of heat flow instantiated on the network.

As such, of the types of traffic - used goods, money, gossip, e-mail, attitudes, infection, and packages - described in \cite{Borgatti}, the spread of {\em attitudes} is modeled by this case.  However, the diffusion mechanism is different than parallel duplication (used in, for example, models of gossip propagation)- instead of a node exerting influence over its neighbors by duplication of its attitude, there is merely a nudging of the attitudes of neighbors which may be either tempered or enhanced by the attitudes of other neighbors.  Thus, we see application of a notion of centrality based on the graph Laplacian to situations where this type of information diffusion is appropriate.  Two of our main examples below - the network of roll call votes in the House of Representatives and the correlation network of the S\&P 500 network - have substantial aspects of this type of information diffusion.  Indeed, in both cases the behavior of the nodes is, at least in part, influenced by the behavior of the nodes they are connected to via measures of similarity.

\subsubsection*{Opinion dynamics}

As mentioned in the introduction, spectral centralities have a direct connection to the models of De Groot \cite{degroot} and Lehrer \cite{Lehrer} (see also \cite{OSM,OSFM} for a broader framework of consensus in multi-agent systems with similar dynamics).  Those models posit a social network encoded in an adjacency matrix $A$ where the matrix is normalized so that the degree of each node is one.  In that case, $L=I-A$.  If we let $x_i(t)$ be the measure of opinion of node $i$ at time $t$, the model then updates the opinions by the weighted sum of neighboring opinions:
\begin{equation*}
\begin{split}
x_{t+1} &= A x_t\\
&= (I-L)x_t\\
&= x_t - Lx_t
\end{split}
\end{equation*}

Thus, under the further restriction that $A$ has been normalized to unit degree, we have that the heat flow modeled by the Laplacian is precisely this model of opinion dynamics.

We note that some investigations using similar spectral perturbation methods in this direction have already been completed to maximize connectivity \cite{degennaro} or synchronization dynamics \cite{ghosh,xiao}.  With the assumptions above and our terminology, these are investigations related to the $1$-spectral centrality.  Interested readers should also see \cite{patterson} for related work when considering social networks which incorporate interaction dynamics.

\section{Results}

In this section, we discuss several different applications of the node $1$-spectral centrality of real networks and compare them to other centrality statistics - degree, betweenness centrality, random walk centrality, information, and eigenvector centrality.
We begin with three examples of small unweighted, undirected networks and move to more complicated networks. The examples convey a general observation - that for small, undirected and unweighted networks which are not particularly dense, $1$-spectral centrality performs often similarly to existing centrality measures.  We next examine three dense weighted networks - a network associated to roll call votes in the U.S. House of Representatives, an equities network, and a biological network derived from the function of the Immune System.  These analyses provide evidence that for complicated networks, particularly dense weighted networks, $1$-spectral centrality is substantially different than other measures.

Higher spectral centralities ($k>1$) behave completely differently from existing measures. The equities network case provides a good example of the environment where the $k$-spectral centralities can be most useful - a dense, complex network with structure determined at multiple scales by (potentially) multiple processes.

In the first two dense examples, the network edges are given by a similarity measure and, in each case, this results in all-to-all connections.  The last dense example is also almost a complete graph, with connections only distinguished by their weights. Thus, we focus on weighted versions of the centrality measures which then allow sensible comparison to spectral centrality.

As we shall see in the analysis, these examples provide different glimpses into the varying interaction and relationship between the different measures. We argue that spectral centralities provide either a better way of understanding the data or provide a new window of investigation.

\subsection*{Unweighted sparsely connected networks}
To place $1$-spectral centrality in the constellation of centrality measures, we will compute it as well as other measures on three simple networks.  The first is a toy model - a hand made network that possesses features which reflect different aspects of notions of centrality.  The other two examples are well known smaller networks which have been substantially studied in the literature - the Zachary karate club \cite{karate} and a social network of dolphins \cite{dolphin}.
\subsubsection*{Toy model}
The toy model consists of 13 nodes.  There is one ``central node" - a pinch point which gives the only connections between two halves of the network.  In each half, there are six nodes with differing connectivity.  Figure \ref{ZK-fig} (top left) shows the structure of the network as well as a comparison of betweenness centrality and $1$-spectral centrality.  We see visually that the two measures are qualitatively the same.  This is reflected in Table \ref{corrtab1}, which shows that all of the measures are fairly highly correlated.  But, there is a distinction - $1$-spectral centrality, betweenness and random walk centrality are very similar. The remaining three (degree, eigenvector and information centralities) are also very similar but the two groups have weaker cross-correlation.  This can again be seen in the figure by observing the degree which, for example, is small for the most ``between" node in the center but higher for others.
\subsubsection*{Zachary karate club}
\begin{figure}[!ht]
\begin{center}
\subfigure[Toy network]{
\includegraphics[scale=0.1]{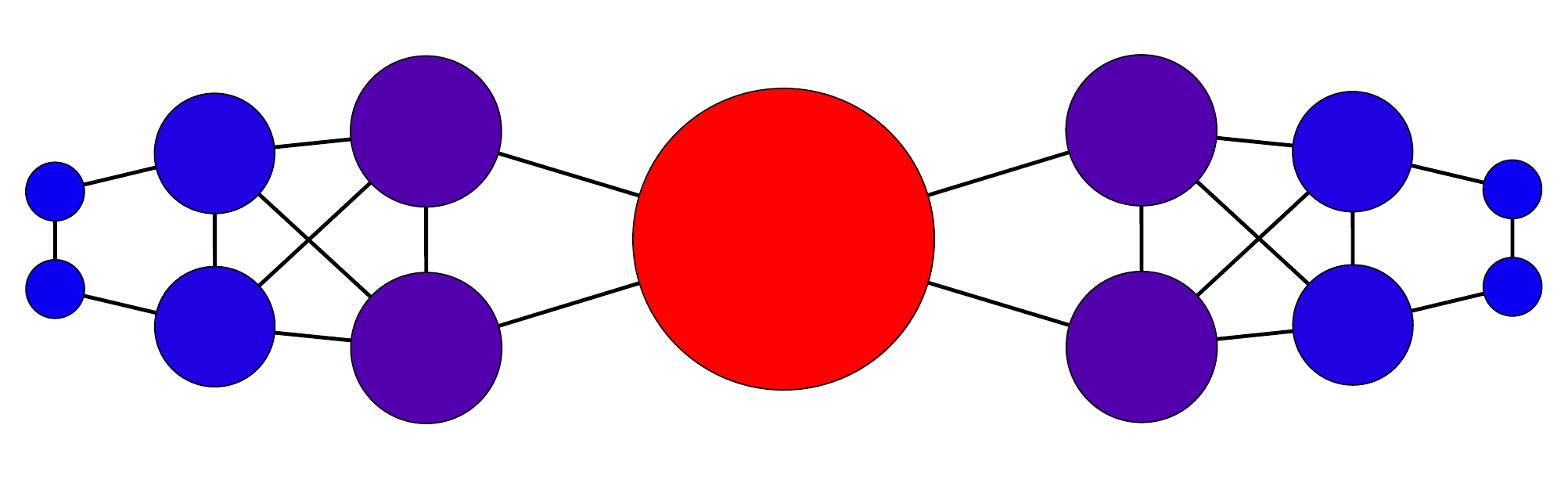}
}

\subfigure[Karate Club]{
\includegraphics[scale=0.1]{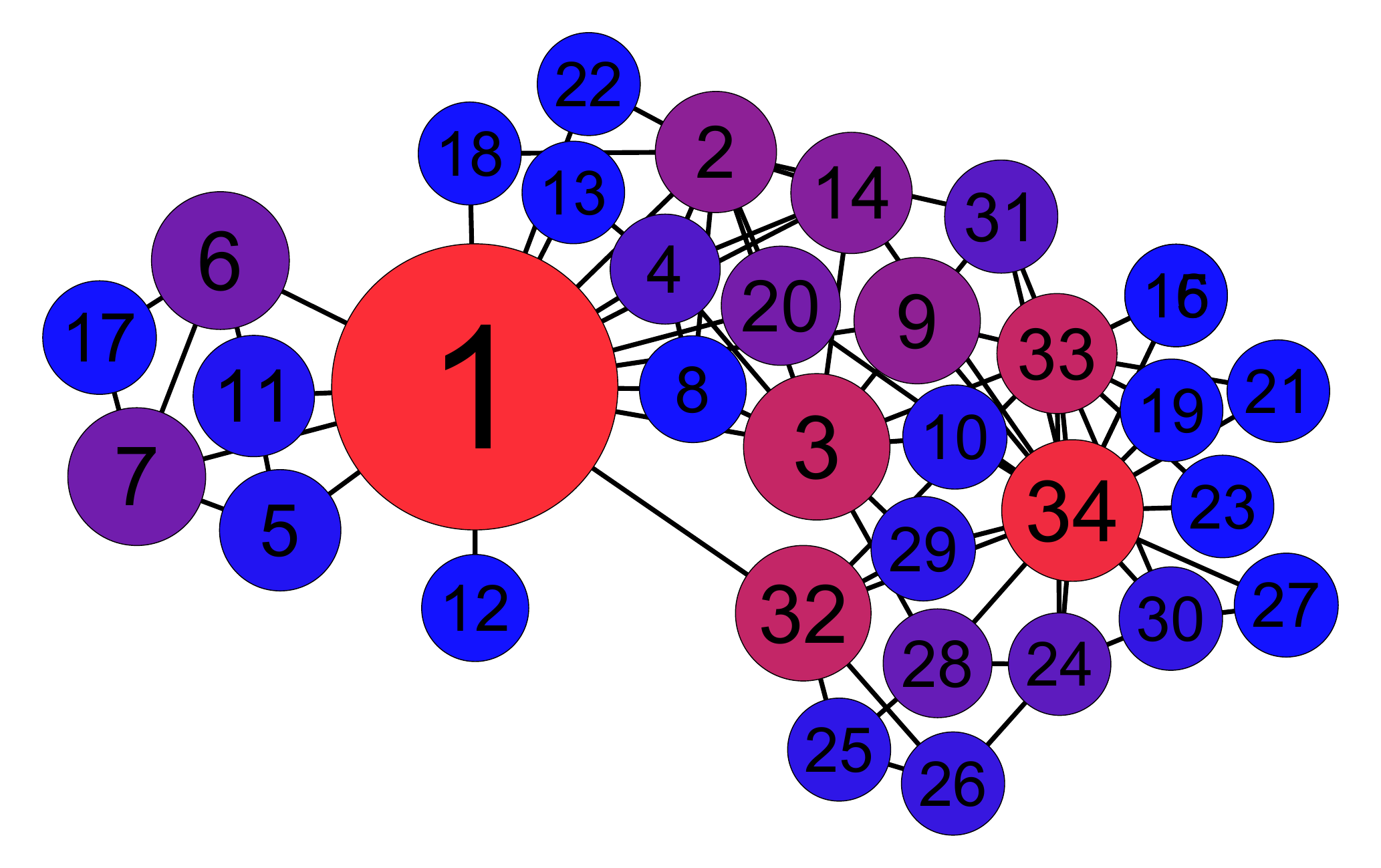}
}
\subfigure[Spectral v. Betweenness Centrality]{
\includegraphics[scale=0.1]{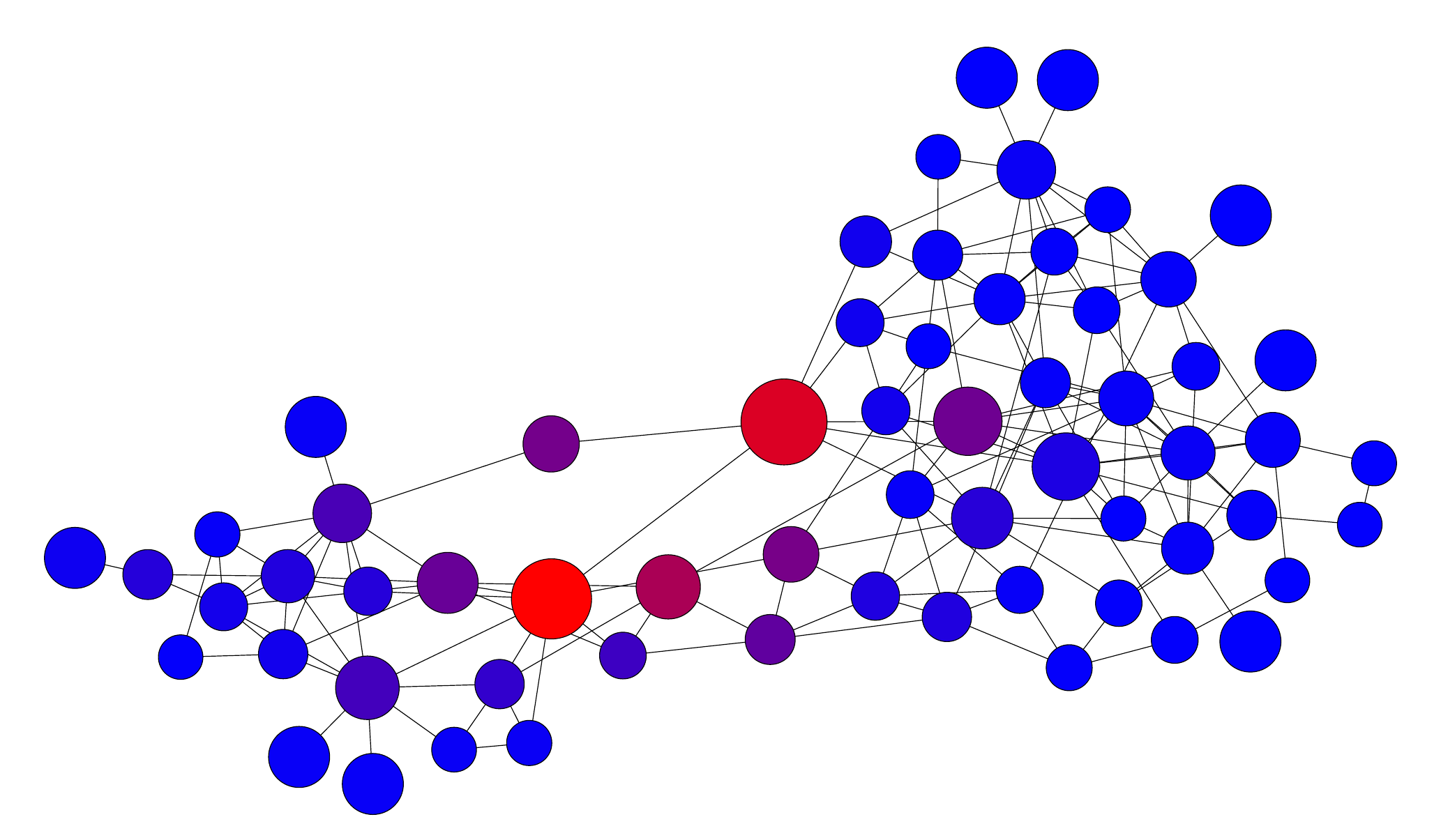}
}

\subfigure[Spectral v. Eigenvector centrality]{
\includegraphics[scale=0.1]{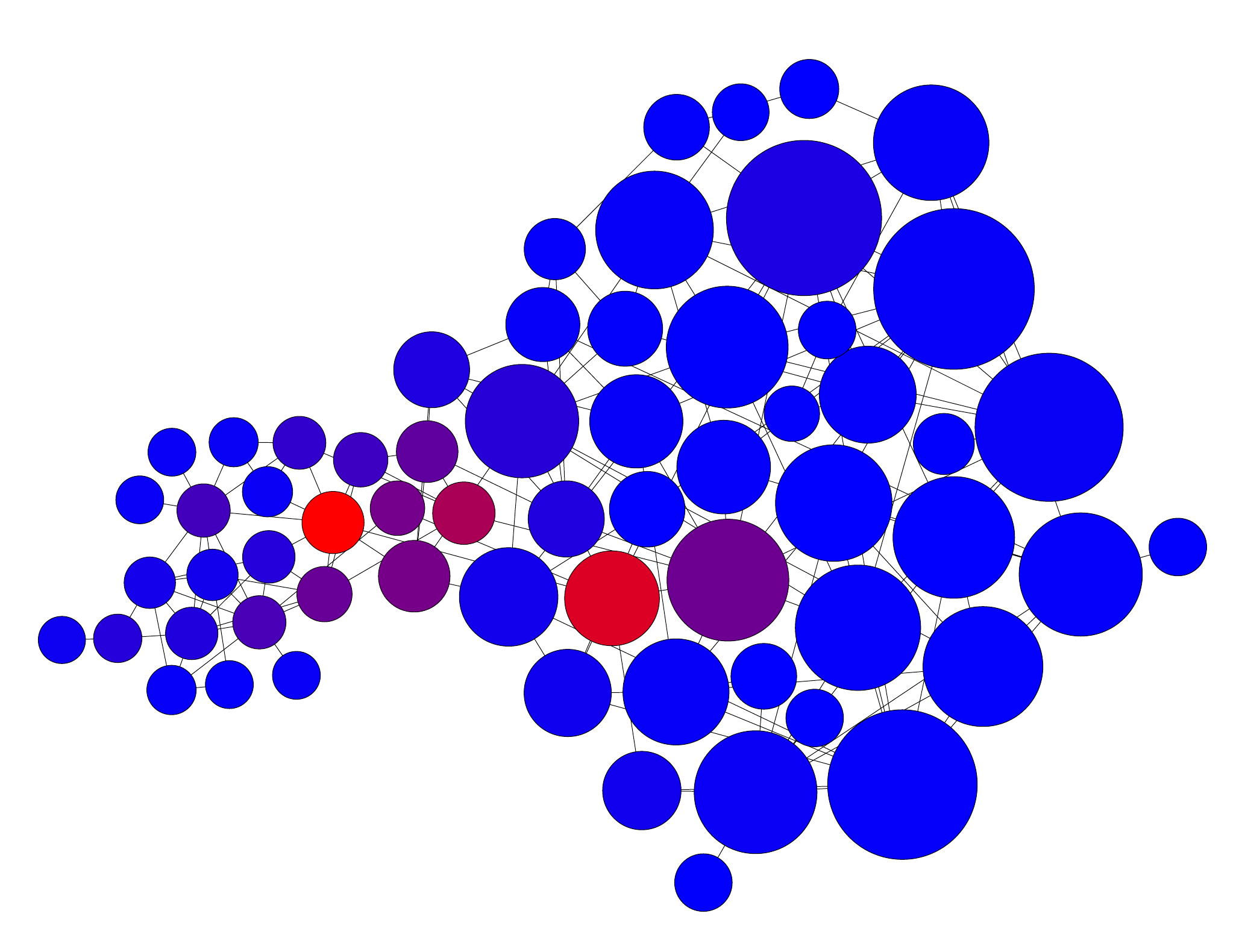}
}
\end{center}
\caption{{\bf Comparison of $1$-spectral centrality and betweenness centrality in the Toy Network, the Zachary Karate club, and the dolphin social network.}  In (a) and (b), size of the nodes is proportional to the $1$-spectral centrality while color (red is high, blue is low) is reflective of the betweenness centrality.  In (c) and (d), color indicates $1$-spectral centrality (red is high, blue is low).  In (c), size is proportional to betweenness and in (d) to eigenvector centrality. }
\label{ZK-fig}
\end{figure}

The Zachary karate club is a social network of 34 nodes, individuals in the karate club, with edges given by social ties.  The adjacency matrix for this network is unweighted and undirected.  Computing degree, betweenness centrality, eigenvector centrality, random walk centrality, information centrality and $1$-spectral centrality, we see that (a general fact that has been previously observed) degree is closely related to all centralities, but that the centralities are differently related to one another.  The results are summarized in Table \ref{corrtab1}, which shows the correlations between the various computations.
Figure \ref{ZK-fig} (top right) illustrates the difference between betweenness centrality and $1$-spectral centrality.  Note that both measures give roughly the same results, but with different emphasis.  This is illustrated by comparing the two for nodes 1 and 34.  Node 1 has the highest centrality in both measures while node 34 has very high betweenness but only moderately high $1$-spectral centrality.  This is reflective of the differences in the ideas behind the two measures.  Node 1 clearly bridges two parts of the network, either from the perspective of finding shortest paths or from the perspective of disconnecting the network (or making it more cohesively connected).  On the other hand, node 34 plays a strong role in the configuration of shortest paths but its removal would not harm the connectivity of the graph as much as some other nodes.

\subsubsection*{Dolphin Network}
This data set consists of 64 nodes (dolphins) and edges between them representing a measure of social ties - in this case, affinity is measured via sustained proximity.  Like the Zachary Karate Club network, the dolphin network is undirected and unweighted.  The correlations of the various centrality measures are again summarized in Table \ref{corrtab1}.  These computations have some similarities with those for the karate club - degree is highly correlated with betweenness, eigenvector, information and random walk centralities.  However, the correlations are generally not quite as strong and $1$-spectral centrality is more weakly correlated with degree.  Moreover, there is a small negative correlation between $1$-spectral and eigenvector centrality. Thus, these two networks show some of the subtleties in the relationships between the various measures.

To give a sense of these relationships, in Figure \ref{ZK-fig} c) and d) we show two views of the network.  In both views, color represents spectral centrality with red indicating high and blue indicating low spectral centralities.  In c), the size of nodes is proportional to betweenness centrality while in d), size is proportional to eigenvector centrality. We see that as betweenness and $1$-spectral centralities are highly correlated, the give qualitatively similar results.  In particular, two dolphins - Beescratch and SN100 - have the highest scores in both measures.  In contrast, the eigenvector centrality is quite different from the $1$-spectral centrality.

\begin{figure}[!ht]
\begin{center}
\includegraphics[scale=0.45]{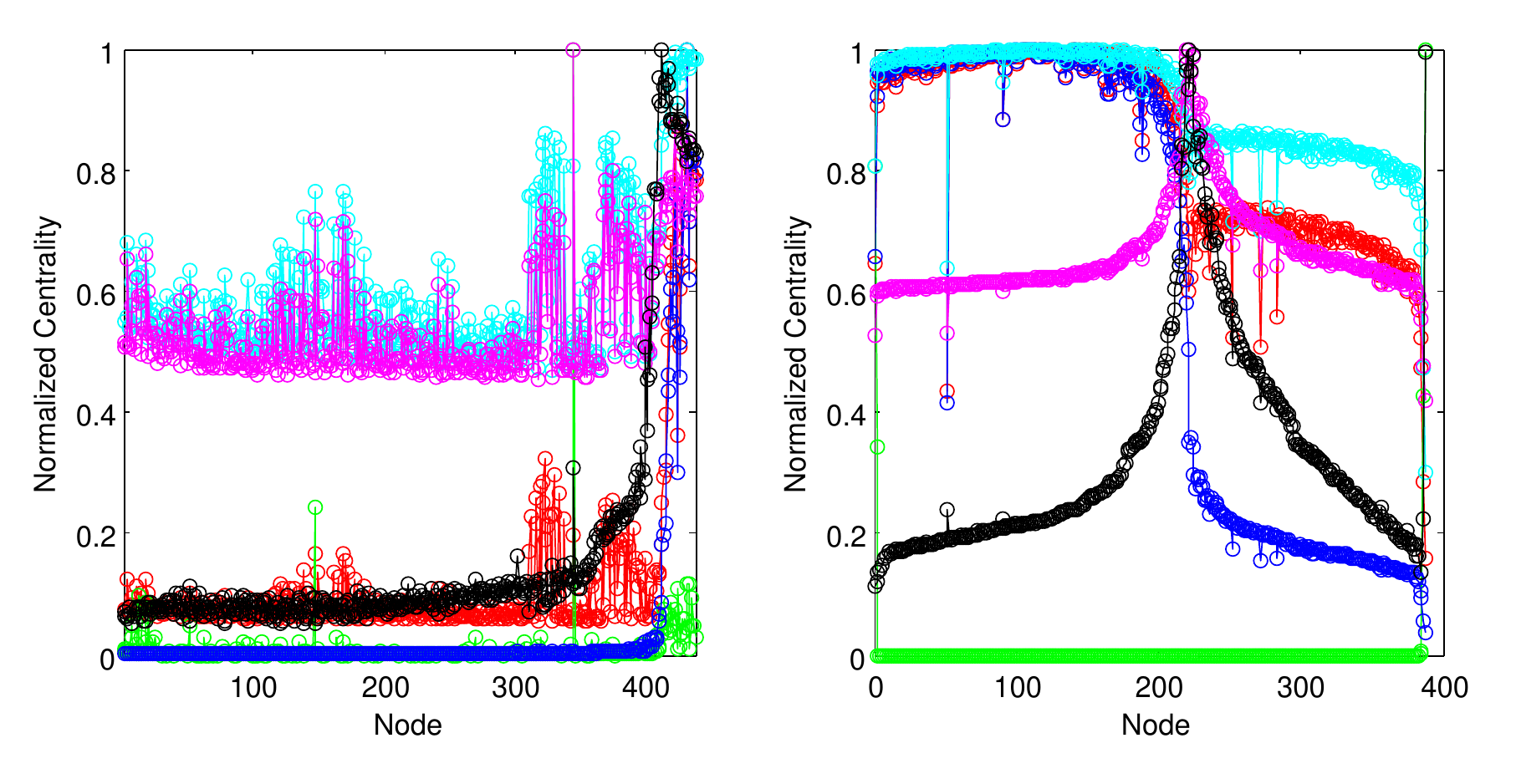}
\end{center}
\caption{{\bf Comparison of the centrality statistics for two dense, weighted networks.} Left: $111\text{th}$ U.S. House roll call network Right: S\&P 500 equities network.  In each case, the x-axis is ordered by the Fiedler vector values of the associated graph Laplacian.  Color indicates the centrality used: red = degree, green = betweenness, blue = eigenvector, cyan = information, magenta = random walk, black = spectral.}
\label{stat-comp}
\end{figure}

\subsubsection*{Roll call votes in the $111^{\text{th}}$ U.S. House of Representatives}

For this example, we use roll call data collected by J. Lewis and K. Poole, available at \cite{voteview}.  The data consist of all roll call votes for the $111^{\text{th}}$ U.S. House of Representatives.  For each member of the House, each of their votes is recorded as a yes ($1$), no ($-1$), or abstain/not present ($0$).  We remove any legislator who did not vote in at least a third of the time.  Then, omitting the missing votes, we compute the percent of votes that pairs of legislators have in common.  The resulting matrix, $P$, has values between $0$ and $1$ and encodes, as a weighted undirected graph, the aggregate voting profiles.  We note that this graph is dense - most legislators vote on most bills, so all legislators have some relation to one another.  To create a weighted adjacency matrix, we use a scaling of the data given by
\begin{equation*}
A_{ij}=
\begin{cases}
e^{-\frac{(1-P_{ij})^2}{\sigma^2}}\;\;\text{for $i\neq j$}\\
0 \;\;\text{for $i=j$}
\end{cases}
\end{equation*}
where $\sigma = 0.25$.  This value of $\sigma$ was selected to focus the analysis on the relevant aspects of the data.  It represents a scaling of the raw percentages we initially computed by accounting for the fact that almost every pair of representatives has at least half of their votes in common due to a host of noncontroversial bills.  To ensure no self loops, we remove the diagonal entries.

\begin{figure}
\begin{center}
\subfigure{
\includegraphics[scale=0.4]{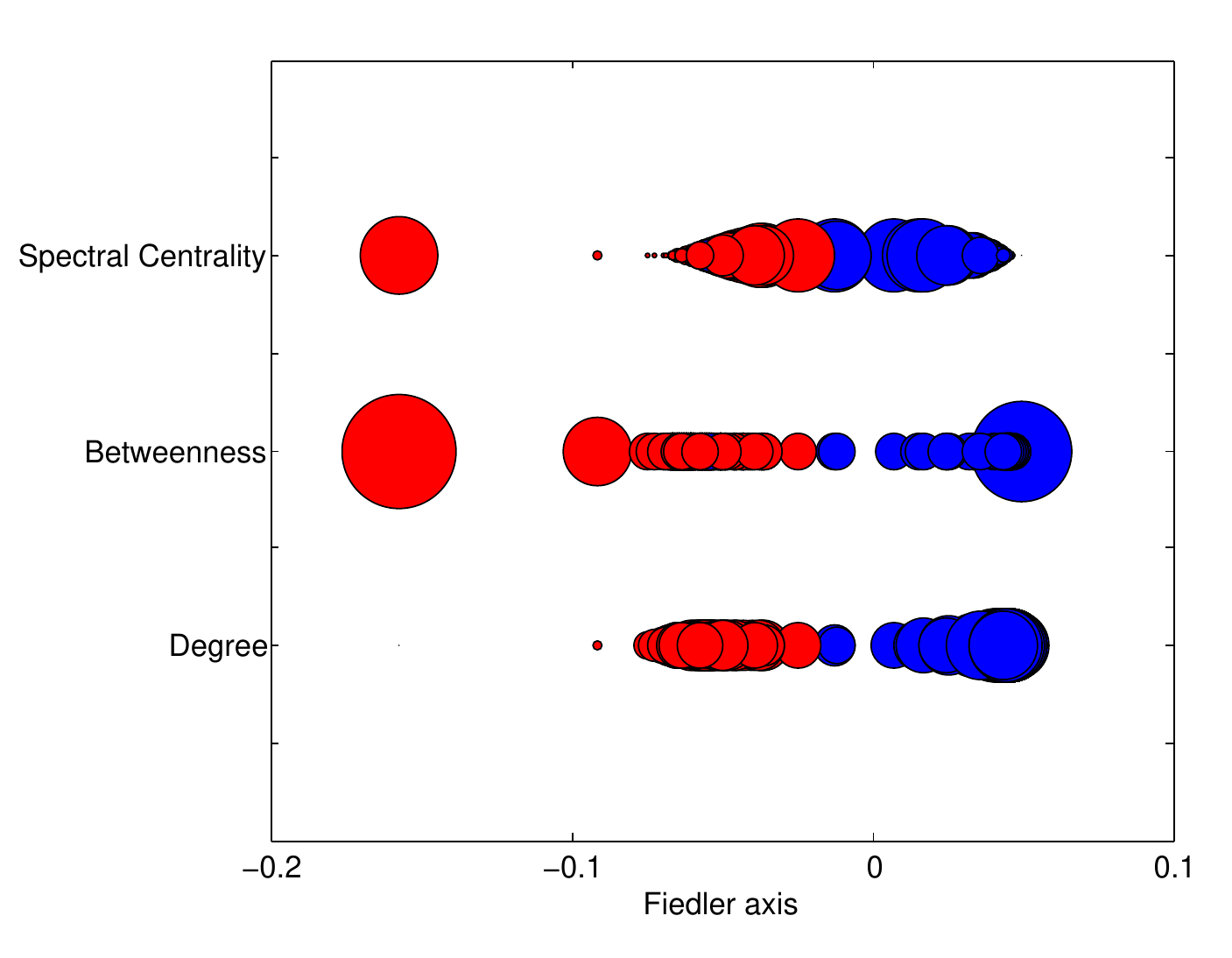}
}
\subfigure{
\includegraphics[scale=0.3]{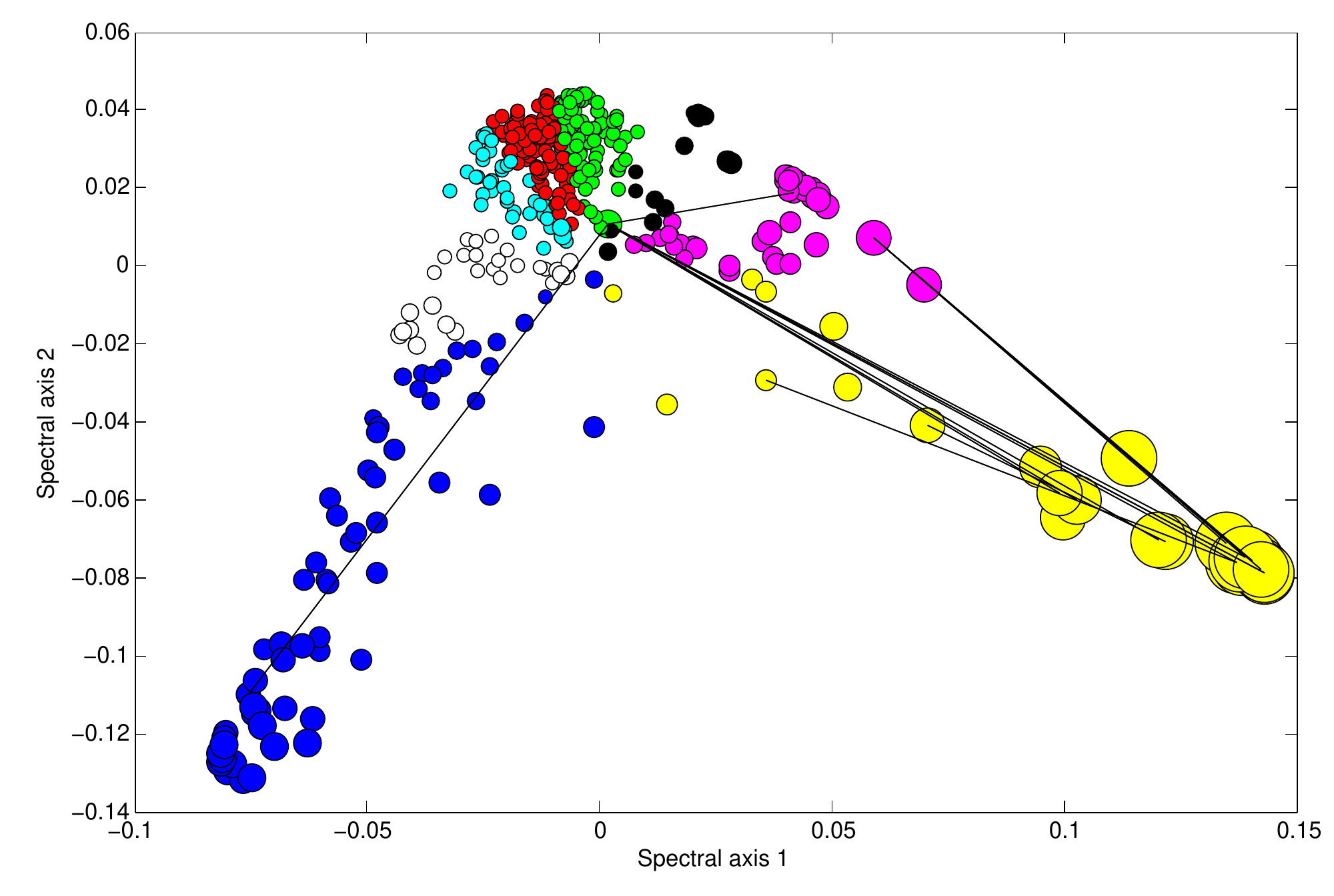}
}
\end{center}
\caption{{\bf $1$-spectral centralities in example dense weighted networks.} Top: A comparison of degree, betweenness and spectral centralities of the roll call network of the $111^\text{th}$ U.S. House of representatives.  Color denotes party affiliation (red=Republican, blue=Democrat) while size is proportional to the relevant statistic.  Nodes are ordered by the Fiedler vector.  Bottom: The S\&P 500 network given by its two dimensional spectral embedding where node size is proportional to spectral centrality.  Coloring is given by an 8-cluster spectral clustering using two eigenvectors, the edges are the 30 edges with the highest spectral centrality.}
\label{fig-House}
\end{figure}

Table \ref{corrtab2} shows the correlations between the various centralities.  In contrast to the previous cases, we see a marked difference between most traditional centralities and $1$-spectral centrality.  Degree, eigenvector, and information centrality are all highly correlated to one another and are {\em negatively} correlated with $1$-spectral centrality.  Betweenness centrality is different from all others, exhibiting only a weak negative correlation with other centralities.  Random walk centrality shows almost a perfect correlation with $1$-spectral centrality.  Figure \ref{fig-House} (a) shows a graphical comparison of degree, betweenness, and $1$-spectral centrality.

This example begins to show one of the positive aspects of $1$-spectral centrality.  Analyses in political science (see \cite{PS1,PS2,PS3}) have shown that recent U.S. Legislative bodies are basically unidimensional with respect to the structure encoded in the roll call votes when analyzed with spatial models.  This structure corresponds to our intuitive identification by party and ideology.  In that sense, members of the House that would be most ``central'' are ones whose ideological preferences lie somewhere between the median preferences of the two major parties.  From Figure \ref{fig-House}, we see that the traditional centrality measures are not overly informative from this point of view - they do not reflect this intuitive notion, instead assigning higher centrality scores to more peripheral members (the most peripheral, all the way to the left, is Rep. Ron Paul (R-TX)).  $1$-spectral centrality, on the other hand, better reflects the intuitive notion of centrality in terms of ideology.

In this example, the $2$-spectral centrality does not contain much useful information.  The $2$-centrality score for one node is very high (roughly $35$) while the rest are between zero and one.  This is possibly a reflection of the basic unidimensionality of the data at this scale - the Fiedler data captures the majority of the information, leaving little to be encoded in the rest of the eigendata. Another indication of the lack of information in the higher spectral centralities is the substantial similarity between the $1$-spectral centrality and the random walk centrality - both of these measure aspects of moving randomly throughout the network, and their similarity indicates that higher order spectral data does not contribute significantly to the diffusion process.

\subsubsection*{Equities network}

In this example, we begin with raw data given daily close prices over a three year period (7/2003-12/2006) of the equities in the S\&P 500 index.  We perform some preprocessing - converting the prices to percent changes from day to day and detrending the results.  We then take the pairwise correlations,$C_{ij}$, between these time series and form a similarity matrix by converting to chordal distance, $S_{ij}=\sin(\arccos(C_{ij})/2)$.  We construct a scaled adjacency matrix:
\begin{equation*}
A_{ij}=
\begin{cases}
e^{-\frac{S_{ij}^2}{\sigma^2}}\;\;\text{for $i\neq j$}\\
0 \;\;\text{for $i=j$}
\end{cases}
\end{equation*}
where $\sigma = 0.25$.   This value of $\sigma$ was selected similarly to the example of the roll call network.  Even after cleaning, the data still had substantial average correlation.  This value of $\sigma$ helps to promote significant correlation and devalue average correlation.

Table \ref{corrtab2} shows the correlations between the various centralities: betweenness centrality is again only weakly associated with other centralities, but degree, eigenvector, random walk, information and $1$-spectral centralities are all highly correlated.

In contrast to the roll call network, spectral centrality in the equities network is best interpreted in terms of its definition - what are the nodes which most determine the shape of the network?  In this case, we see two groups with higher than average centrality which correspond to two sectors - the Basic Materials sector and the Technology sector.  Roughly, the same groups are identified by the centralities correlated with $1$-spectral centrality (14 nodes in common in the top-ranking 20 for spectral centrality and degree).  The 12 remaining nodes come from all positions on the list.  
Thus, $1$-spectral centrality is detecting nodes with fairly low scores in other centralities.  This reflects the fact that such nodes may still have a substantial contribution to the shape of the network despite not scoring particularly highly on other measures.  One way to view this is that these nodes have connections which are significant to the solution to the relaxed ratio-cut problem.  Figure \ref{fig-House} illustrates this idea.  In that picture, nodes are placed according to the two dimensional spectral embedding with sizes proportional to their $1$-spectral centrality.  The included edges are the thirty edges with the highest spectral edge centrality.  One can see from this illustration how $1$-spectral centrality reveals structure.  The nodes in the lower right corner, a cluster of Technology equities, are the most central.  If one computes the relaxed ratio cut, it is essentially this cluster which is separated from the rest.  Evidence of this is encoded in the edges - most of these thirty most central edges connect one of these nodes to nodes in other areas of the network.  However, some connect other pieces, showing higher order structure.

\begin{figure}[!ht]
\begin{center}
\includegraphics[scale=0.4]{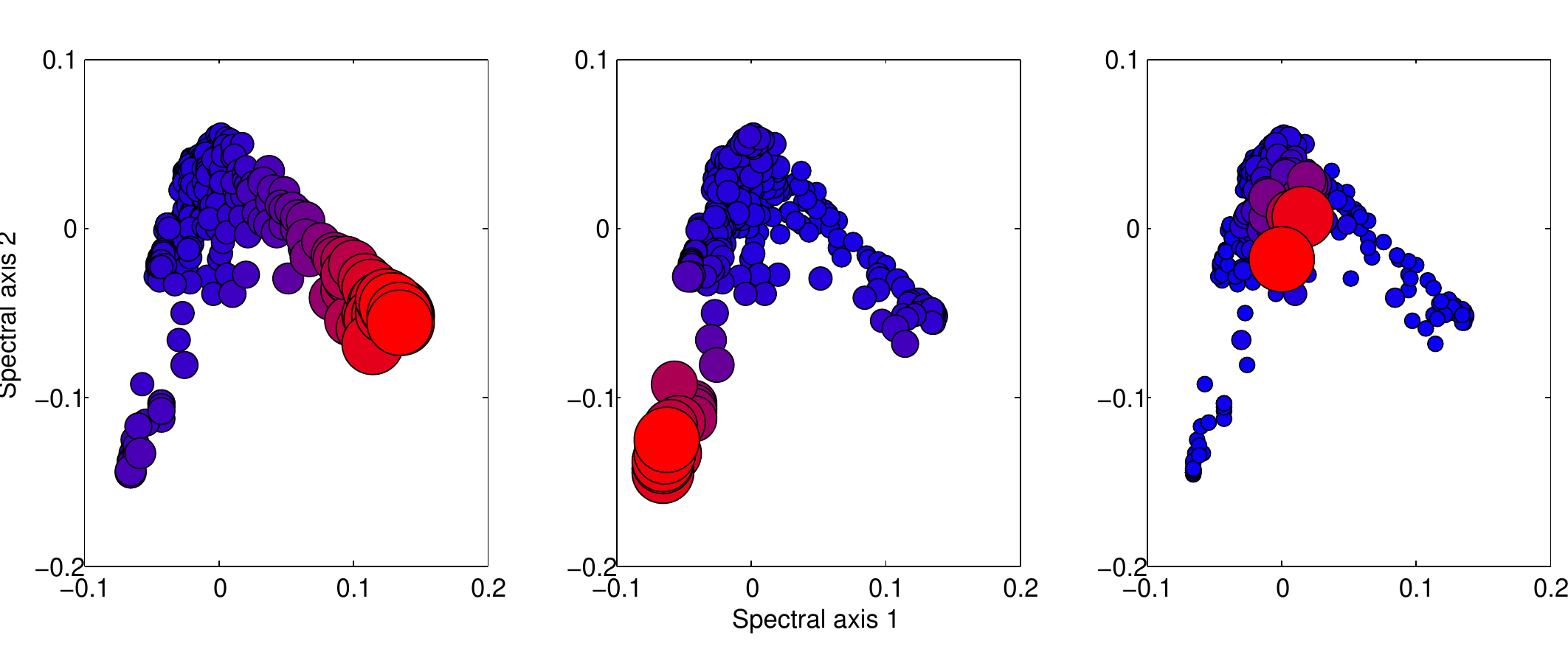}
\end{center}
\caption{{\bf $1$, $2$, and $3$-spectral centralities of the S\&P 500 network.}  Size is proportional to centrality scores. Color also reflects spectral centrality, with red indicating higher scores and blue lower ones.}
\label{highercent}
\end{figure}

Higher order spectral centralities carry interesting and useful information.  Figure \ref{highercent} shows the same spectral embedding of the network as shown in Figure \ref{fig-House} with node sizes proportional to the $1$-spectral centrality (left), $2$-spectral centrality (middle), and $3$-spectral centrality (right).  The colors of the nodes in this figure are also indicative of the $k$-spectral centralities with red indicating higher values and blue indicating lower ones.  These three centralities show three distinct clusters of high centrality nodes.  As discussed above, the cluster in the lower right hand side shows the Technology sector, which have high $1$-spectral centrality values.  The cluster in the lower left hand corner, with high $2$-spectral centrality, are members of the Basic Materials sector.  The cluster emerging from the middle when scaled by the $3$-spectral centrality is a group of equities from multiple sectors, with the majority being from the Healthcare sector.

\begin{figure}[!ht]
\begin{center}
\includegraphics[scale=0.45]{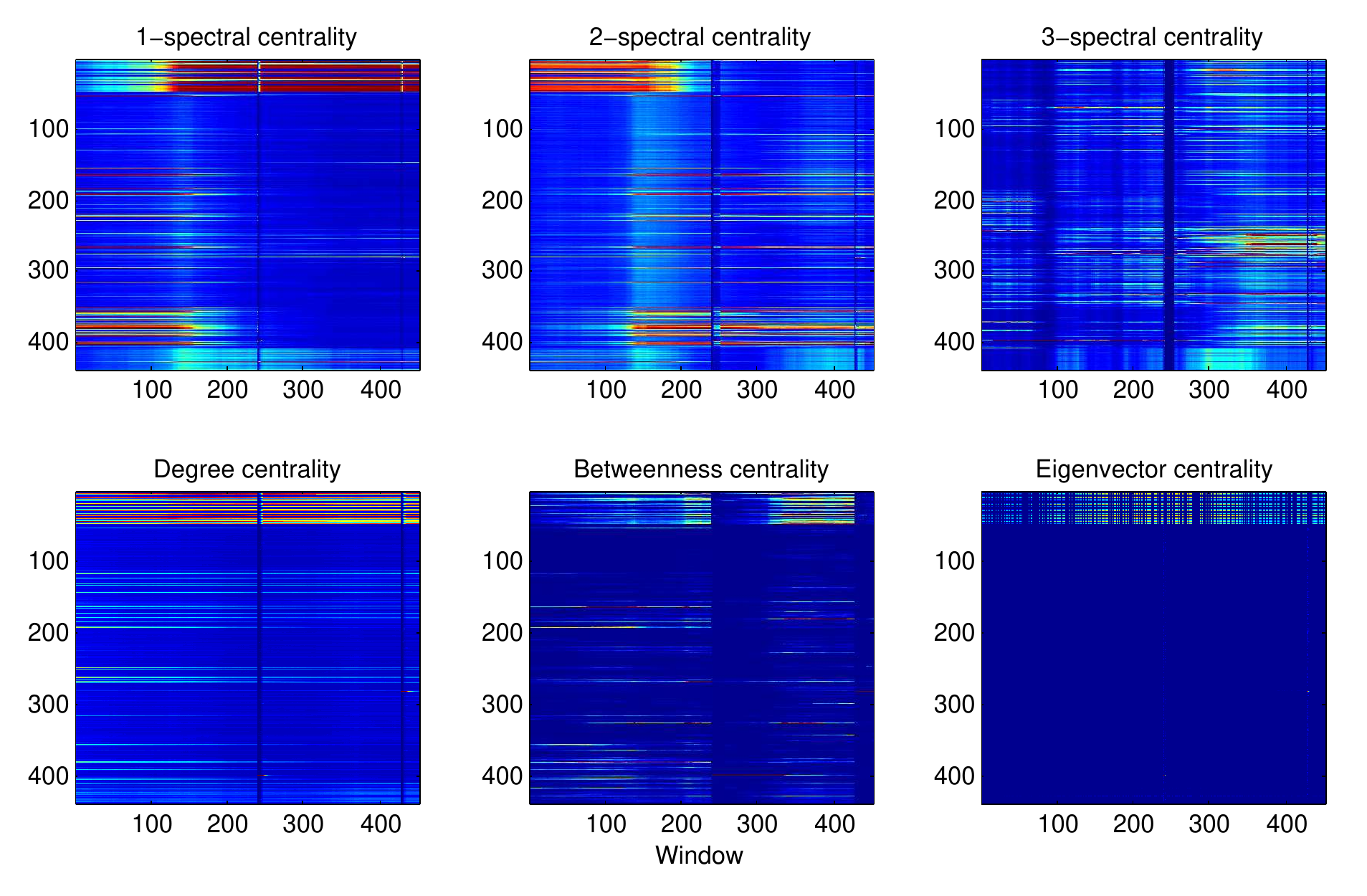}
\end{center}
\caption{{\bf A comparison of spectral centrality and other centralities over windows of the S\&P 500 data.}}
\label{sp-win}
\end{figure}

To better understand how the spectral centralities encode structure, we consider a sequence of windows of the S\&P 500 data.  We compute the matrix $A$ described above for every 450 day window of the data and compute the centralities for each window.  This allows us to see how the centrality of various nodes changes over time.  Figure \ref{sp-win} shows the results for $1$, $2$, and $3$-spectral centrality (top row), degree (bottom left), betweenness (bottom middle), and eigenvector centralities (bottom right).  In all cases, the nodes, as labeled by the y-axis, are ordered by sector identification while the x-axis shows the window index.  For each case, the centralities are normalized to take values in $[0,1]$ so as to make the computations comparable.

We see that in the case of $1$-spectral centrality, there is a clear transition between one group, which is high for roughly the first 150 windows, and another which is high for the rest of the windows.  The majority of equities in the first group are members of the Technology sector while the second group is dominated by the Basic Materials sector.  $1$-Spectral centrality shows an aspect of the effect of the business cycle on the equities market - recording a transition from the dominance of Technology stocks to that of Basic Materials stocks.  $2$-Spectral centrality shows roughly the opposite picture, demonstrating how the first two eigenvalue/eigenvector pairs encode this transition.  $3$-Spectral centrality shows a complementary picture, where a different sector (Healthcare) rises in importance in later windows.

In contrast, the same picture for the other centralities miss the dynamic picture entirely.  Degree and eigenvector centralities show prominence only for the Basic Materials sector and with a peak at a different time than that of spectral centrality.  Betweenness again emphasizes Basic Materials and scattered other equities of importants.

Spectral centralities also help us determine the relevance of the results of spectral clustering.  The color of the nodes in Figure \ref{fig-House} is given by cluster membership where we have run the spectral clustering algorithm for five clusters using two eigenvectors. In choosing the parameters for spectral clustering - the number of clusters and number of eigenvectors - one is always faced with the consequences of making poor choices.  Choosing too many clusters leads to a version of overfitting, vectorization, where natural geometric clusters can be artificially subdivided.  In this figure, we can see how the combination of edge and node spectral centrality sheds light on this issue.  Qualitatively, we see that the blue, yellow and purple clusters are significant from the perspective of spectral methods while the white, black, and light blue clusters are more likely due to vectorization.  The green cluster, which sits among the latter group, has relevance, containing nodes and edges with substantial $3$-spectral centrality.  This suggests that spectral centrality measures may be used in conjunction with spectral clustering to optimize parameter choices.

\subsubsection*{Immune System mediator network}

For our last example, we consider a dense weighted network describing the mediators in the human immune cell network \cite{Tieri05}.
From data available in the literature \cite{database}, we construct a bipartite graph between 19 cell types of the Immune System (e.g. T and B lymphocytes, neutrophyls, etc.) and 109 mediators of their mutual interaction (such as cytokines, chemokines, etc.).  This graph is then collapsed to directed cell-cell interaction network. While the network is almost completely connected, the link weights - indicating the number of mediators for each pairs of cell types - carry the structure.

In \cite{Tieri05} an efficiency measure \cite{Infc} was used to characterize the mediator relevance inside the network and quantify the relevance of each mediator. This was achieved by selective removal of the mediator from the original bipartite graph, collapsing the network and comparing the efficiency of the resulting network to that of the original.

For our application, we compute the 1-spectral centrality of the symmetrized cell-cell network, $C_0$, via a deformation associated with each mediator.  We let $B_i=C_0-C_i$ where $C_i$ is the symmetrized cell-cell network after the $i^\text{th}$ mediator has been removed from the bipartite network and the bipartite network has been collapsed.  We then compute the 1-spectral centrality using equation \eqref{speccentgen}.

\begin{figure}[!ht]
\begin{center}
\includegraphics[scale=0.2]{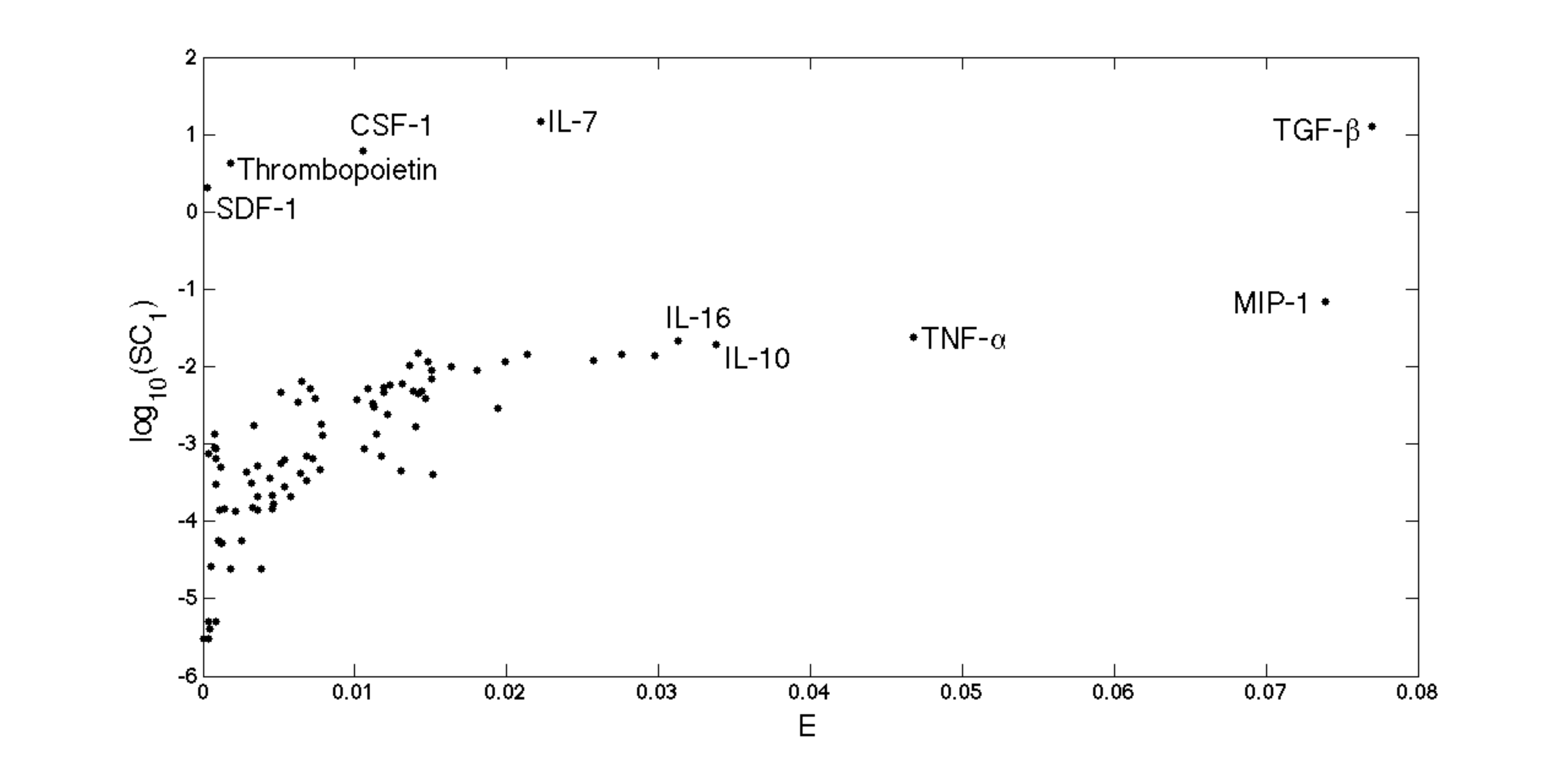}
\end{center}
\caption{{\bf Comparison between Efficiency and 1-Spectral Centrality for a Immune System mediator network.}}
\label{cyt_fig}
\end{figure}

Some of the mediators receive a high ranking from both centrality measures (e.g. TGF-$\beta$, a fundamental anti-infiammatory citokyne) except for a group of mediators on the top left of the figure (CSF-1, SDF-1, Thrombopoietin and IL-7) which have high 1-spectral centrality but low Efficiency.
In the previous analysis, top-ranking mediators were related to innate response in the Immune system, an ancient and highly conserved mechanism that involves among the oldest elements of the mediator network. The efficiency centrality measure thus seems to have selected the mediators that reflect the Immune System evolution, in which the oldest elements should be also the most connected ones (as it happens in a preferential attachment-like model of network growth).
Interestingly, the most central mediators obtained by 1-SC have a completely different role, since they are known from literature to be jointly involved in cell maturation and differentiation \cite{hematopoiesis}, a mechanism known as hematopoiesis, fundamental for the maintenance of the whole Immune System.  Even if we have a large overlap with the previous analysis (the two centrality measures look nonlinearly correlated), the information about mediator relevance from the two measures is not the same.
The relevance of the mediators in the network as attributed by 1-SC seems to be less dependent on network evolution issues, and more related to biological and functional aspects.

This example demonstrates the use of general set deformations for spectral centrality measurements.
In this case, there was a natural deformation associated with each mediator.
However, because the 1-spectral centrality is related to a subset of the network determined by an underlying bipartite network, it is inappropriate to compare them to other centrality measures of $C_0$.


\section{Discussion}
We introduce a new notion of centrality for graphs, $k$-spectral centrality, based on measuring the effects of deformations (i.e. node, edge or network subset removal) of the graph Laplacian on its eigenvalues.  

The $k$-spectral centralities reveal information concerning the importance of different parts of the graph with respect to its geometry.  For example, $1$-spectral centrality identifies features relevant to the algebraic connectivity.  Higher spectral centralities yield information relevant to the spectral embedding of the graph as well as to spectral clustering.  We see that for small, sparse, unweighted, symmetric networks, spectral centrality behaves similarly to other standard centralities.  However, for larger, denser, weighted symmetric networks, spectral centrality provides a distinct, new, and potentially useful tool in network analysis.

We see an initial setting in which the $k$-spectral centralities are appropriate - consensus formation.  For networks which reflect the substrate on which opinion propagates and consensus does or does not form, the $k$-spectral centralities provide relevant summary statistics of importance.  Basic models of consensus formation posit opinion evolution via dynamics governed by the graph Laplacian.  Thus, by construction, the $k$-spectral centralities measure importance with respect to those opinion dynamics.  Our first two dense weighted network examples can be interpreted through the lens of consensus formation.  In each case the behavior of the nodes is correlated (at different strengths) with that of other nodes.  Background opinion formation dynamics - opinions about bills in the U. S. House of Representatives and opinions of traders about the similarities of different equities - create a link between the correlation and consensus dynamics.  In the example of the U.S. House of Representatives, if we assume that representatives are influenced by other representatives with strengths associated with their social/professional ties, this network is then well modeled by the consensus formation dynamics.  For the equities market, we assume a hidden network of traders who form opinions about the equities they trade.  The opinions translate into close prices for equities as the traders buy and sell their various holdings.  Thus, the time series of close prices is a reflection of the consensus dynamics of the traders and exhibits its own dynamics as a proxy for the aggregated dynamics of the traders.  The equities network we construct is therefore a proxy for a consensus network of unseen actors.

In this context, we see that the collection of $k$-spectral centralities deliver better measures of centrality than existing measures.  In the roll call network, the $1$-spectral centrality reflects our common notion of ideological centrality better than all other measures except random walk centrality.  But, the lack of information in the higher centralities confirms that the information in the $1$-spectral centrality is essentially complete.  In contrast, the first three spectral centralities for the S\&P 500 network all carry significant information which show three different groups of important nodes and identifies the scale at which they are important.  In our experiment with centrality measures over a windowed version of the S\&P 500 network, we see further evidence that the spectral centralities outperform other measures in this application.  Indeed, the $1$-spectral centrality is the only measure to capture the evolution of importance and centrality as the network changes over time.

The third example of a dense weighted network, the Immune System network, provides an example where it has already been observed that common centrality measures fail to provide a clear ranking of mediator relevance \cite{Tieri05}, efficiency was therefore used to extract a measure of relevance for mediators in this network.  1-spectral centrality also provides a measurement of relevance for mediators, but it provides a different ranking.  While some mediators involved in anti-inflammatory responses had both high efficiency and 1-spectral centrality, the majority of mediators with highest 1-spectral centrality also had low efficiency.  These mediators are involved in cell maturation and differentiation (hematopoiesis) while the top-ranking mediators by efficiency are more related to the biological process of innate response of the Immune System. While our centrality measure highlights mediators related to fundamental system functions (its maintenance by means of cell maturation and differentiation), efficiency ranking in this case seems to depend on the process of network formation, for which the oldest network elements are also the most central. This example provides evidence that 1-spectral centrality is a useful tool in identifying functional structure in biological networks which complements other existing measures.

\section{Acknowledgments}
D.R. would like to thank Stefano Salvioli for useful discussion on Immune System Network.

\bibliography{plos-spec-cent}

\section*{Tables}


\begin{table}[!ht]
\caption{
\bf{Correlations between centralities for unweighted example networks}}
\small
\begin{tabular}{|c|c|c|c|c|c|c||c|c|c|c|c|c||}
  \hline
   & \multicolumn{6}{c||}{Toy Network} &\multicolumn{6}{c||}{Zachary Karate Club} \\
\hline
   & D& B & E& RW &I & S& D& B & E& RW & I & S\\\hline
  D & 1 & 0.70 & 0.97&0.74 & 0.92 & 0.48 & 1 & 0.92& 0.92 & 0.98 & 0.89 & 0.70   \\ \hline
  B &   & 1 & 0.81 & 0.99 & 0.91 & 0.96 &   & 1 & 0.80 & 0.96 & 0.74 & 0.87  \\ \hline
  E &   &   & 1 & 0.84 & 0.98 &0.64  &   &   & 1 & 0.89 &0.96 & 0.58 \\ \hline
  RW & & & &1 & 0.94 & 0.94 &  & & & 1 & 0.87 & 0.81 \\ \hline
  I & & & & & 1 & 0.79 & & & & & 1 & 0.55 \\ \hline
  S &  & & &  &   & 1&   & &  &  & & 1\\  \hline
 & \multicolumn{6}{c||}{Dolphin Network} &\multicolumn{6}{c||}{} \\
\hline  
  D & 1 & 0.59 & 0.72 & 0.81 & 0.89 & 0.22 & & & &  &   &  \\ \hline
  B &  & 1 & 0.28 & 0.89 & 0.55 & 0.82 & & & &  &   &  \\ \hline
  E &  &   & 1 & 0.39 & 0.74 & -0.13   & & & &  &   &  \\ \hline
  RW &   &   &   & 1 & 0.79 & 0.72 & & & &  &   &  \\ \hline
  I  &  & & & & 1 & 0.28 & & & &  &   &  \\ \hline
  S &  & & &  &   & 1 &  & & &  &   &  \\ \hline
  
  \end{tabular}

\begin{flushleft}
Correlations between various centralities for the unweighted example networks we consider.  The rows and columns are labeled as follows:  D = degree, B = Betweenness Centrality, E = Eigenvalue Centrality, RW = Random Walk Centrality, I = Information centrality and S = Spectral Centrality.
\end{flushleft}
\label{corrtab1}
\end{table}

\begin{table}[!ht]
\caption{
\bf{Correlations between centralities for dense weighted example networks}}
\small
\begin{tabular}{|c|c|c|c|c|c|c||c|c|c|c|c|c|}
  \hline
     &  \multicolumn{6}{c||}{Equities network} & \multicolumn{6}{c|}{Roll call network} \\
\hline
   & D& B & E& RW & I& S& D& B & E& RW & I &S\\\hline
  D & 1 & 0.26 & 0.95&0.78 & 0.86 & 0.78 &   1 & -0.24 & 0.96 & -0.25 &0.99 & -0.40 \\ \hline
  B &   & 1 & 0.23 & 0.41 & 0.28 & 0.24 &     & 1 & -0.07 & -0.21 & -0.34 & 0.08 \\ \hline
  E &   &   & 1 & 0.59 & 0.71 & 0.80     &   &   & 1 & -0.30 & 0.91 &-0.37   \\ \hline
  RW & & & &1 & 0.96& 0.66 &   & & & 1 & -0.17 & 0.92 \\ \hline
  I & & & & & 1 & 0.70 & & & & & 1 & -0.36  \\ \hline
  S &   & &  &  & & 1&    & &  &  & & 1 \\  \hline
\end{tabular}
\begin{flushleft}
Correlations between various centralities for the weighted example networks we consider.  The rows and columns are labeled as follows:  D = degree, B = Betweenness Centrality, E = Eigenvalue Centrality, RW = Random Walk Centrality, I = Information centrality and S = Spectral Centrality.
\end{flushleft}
\label{corrtab2}
\end{table}

\end{document}